\documentclass[preprints,article,accept,moreauthors,dvi2pdf,10pt,a4paper]{Definitions/mdpi}

\firstpage{1} 
\pubvolume{xx}
\issuenum{1}
\articlenumber{5}
\pubyear{2019}
\copyrightyear{2019}
\history{Received: date; Accepted: date; Published: date}

\pdfoutput=1

\usepackage{subcaption}
\usepackage[export]{adjustbox}

\Title{Integrated Information Theory and Isomorphic Feed-Forward Philosophical Zombies}

\Author{Jake R. Hanson $^{1,2}$ and Sara I. Walker $^{1,2,3,*}$\orcidB{}}

\AuthorNames{Jake R. Hanson and Sara I. Walker}

\address{%
$^{1}$ \quad School of Earth and Space Exploration, Arizona State University, Tempe, AZ, USA\\
$^{2}$ \quad Beyond Center for Fundamental Concepts in Science, Arizona State
University, Tempe, AZ, USA\\
$^{3}$ \quad ASU–SFI Center for Biosocial Complex Systems, Arizona State
University, Tempe, AZ, USA}

\corres{Correspondence: jake.hanson@asu.edu}

\abstract{
Any theory amenable to scientific inquiry must have testable consequences. This minimal criterion is uniquely challenging for the study of consciousness, as we do not know if it is possible to confirm via observation from the outside whether or not a physical system knows what it feels like to have an inside - a challenge referred to as the "hard problem'' of consciousness. To arrive at a theory of consciousness, the hard problem has motivated development of phenomenological approaches that adopt assumptions of what properties consciousness has based on first-hand experience and, from these, derive the physical processes that give rise to these properties. A leading theory adopting this approach is Integrated Information Theory (IIT), which assumes our subjective experience is a ``unified whole”, subsequently yielding a requirement for physical feedback as a necessary condition for consciousness. Here, we develop a mathematical framework to assess the validity of this assumption by testing it in the context of \textit{isomorphic} physical systems with and without feedback. The isomorphism allows us to isolate changes in $\Phi$ without affecting the size or functionality of the original system. Indeed, we show that the only mathematical difference between a "conscious" system with $\Phi>0$ and an isomorphic "philosophical zombie" with $\Phi=0$ is a permutation of the binary labels used to internally represent functional states. This implies $\Phi$ is sensitive to functionally arbitrary aspects of a particular labeling scheme, with no clear justification in terms of phenomenological differences. In light of this, we argue any quantitative theory of consciousness, including IIT, should be invariant under isomorphisms if it is to avoid the existence of isomorphic philosophical zombies and the epistemological problems they pose.
}


\begin{document}

\section{Introduction} \label{introduction}
The scientific study of consciousness walks a fine line between physics and metaphysics. On the one hand, there are observable consequences to what we intuitively describe as consciousness. Sleep, for example, is an outward behavior that is uncontroversially associated with a lower overall level of consciousness. Similarly, scientists can decipher what is intrinsically experienced when humans are conscious via verbal reports or other outward signs of awareness. By studying the physiology of the brain during these specific behaviors, scientists can build "neuronal correlates of consciousness" (NCCs), which specify where in the brain conscious experience is generated and what physiological processes correlate with it \cite{rees2002neural}. On the other hand, NCCs cannot be used to explain \textit{why} we are conscious or to predict whether or not another system demonstrating similar properties to NCCs is conscious. Indeed, NCCs can only tell us the physiological processes that correlate with what are assumed to be the functional consequences of consciousness and, in principle, may not actually correspond to a measurement of what it is like to have subjective experience  \cite{Chalmers1995}. In other words, we can objectively measure behaviors we assume accurately reflect consciousness but, currently, there exist no scientific tools permitting testing our assumptions. As a result, we struggle to differentiate whether a system is truly conscious or is instead simply going through the motions and giving outward signs of, or even actively reporting, an internal experience that does not exist (e.g. \cite{searle1980minds}).

This is the "hard problem" of consciousness \cite{Chalmers1995} and it is what differentiates the study of consciousness from all other scientific endeavors. Since consciousness is subjective (by definition), there is no objective way to prove whether or not a system experiences it. Addressing the hard problem, therefore, necessitates an inversion of the approach underlying NCCs: rather than starting with observables and deducing consciousness, one must start with consciousness and deduce observables. This has motivated theorists to develop phenomenological approaches that adopt rigorous assumptions of what properties consciousness must include based on human experience, and, from these, "derive" the physical processes that give rise to these properties. The benefit to this approach is not that the hard-problem is avoided, but rather, that the solution appears self-evident given the phenomenological axioms of the theory. In practice, translating from phenomenology to physics is rarely obvious, but the approach remains promising. 

The phenomenological approach to addressing the hard problem of consciousness is exemplified in Integrated Information Theory (IIT) \cite{tononi2008, oizumi2014phenomenology}, a leading theory of consciousness. Indeed, IIT is a leading contender in modern neuroscience precisely because it takes a phenomenological approach and offers a well-motivated solution to the hard problem of consciousness \cite{tononi2016integrated}. Three phenomenological axioms form the backbone of IIT: information, integration, and exclusion. The first, \textit{information}, states that by taking on only one of the many possibilities a conscious experience generates information (in the Shannon sense, {\it e.g.} via a reduction in uncertainty \cite{shannon1948}). The second, \textit{integration}, states each conscious experience is a single "unified whole". And the third, \textit{exclusion}, states conscious experience is exclusive in that each component in a system can take part in at most one conscious experience at a time (simultaneous overlapping experiences are forbidden). Given these three phenomenological axioms, IIT derives a mathematical measure of integrated information - $\Phi$ - that is designed to quantify the extent to which a system is conscious based on the logical architecture (i.e. the "wiring") underlying its internal dynamics.

In constructing $\Phi$ as a phenomenologically-derived measure of consciousness, IIT must assume a connection between its phenomenological axioms and the physical processes that embody these axioms. It is important to emphasize that this assumption is nothing less than a proposed solution to the hard problem of consciousness, as it connects subjective experience (axiomatized as integration, information, and exclusion) and objective (measurable) properties of a physical system. As such, it is possible for one to accept the phenomenological axioms of the theory without accepting $\Phi$ as the correct quantification of these axioms and, indeed IIT has undergone several revisions in an attempt to better reflect the phenomenological axioms in the proposed construction of $\Phi$ \cite{tononi2004information, balduzzi2008integrated,oizumi2014phenomenology}. If, on the other hand, one accepts $\Phi$ as the correct mathematical translation of the theory's phenomenological axioms, no experimental result can disprove the theory because the theory automatically dictates the interpretation of experimental results if the axioms hold. 

The possibility of multiple mathematical interpretations of the same phenomenological axioms implies that, in principle, two competing theories of consciousness can disagree on the results of an experiment even if they accept the same phenomenological axioms - a situation that arises precisely because of the hard problem. Justification for a given phenomenological theory, therefore, must ultimately come from how well the deductions of the theory match our intuitive understanding of what consciousness is, as well as the logical consistency and believability of the underlying assumptions. In this regard, it is important to thoroughly understand any unique or counterintuitive predictions that are deduced, because accepting or rejecting these conclusions is the only way to even approach testing the validity of the theory \cite{godfrey2009}.

In the context of IIT, the implied existence of philosophical zombies is a particularly controversial claim. Philosophical zombies are defined as physical systems that are capable of perfectly emulating the outward behavior of conscious systems but which nonetheless, lack subjective experience. Epistemologically, the existence of philosophical zombies is problematic, as many have argued that it is logically impossible for a scientific theory to justify their existence \cite{Turing1950,harnad1995,doerig2019}. The fact that IIT admits such systems, therefore, poses a serious threat to the validity of the assumptions that form the foundations of the theory and, in particular, IIT's mathematical translation of the integration axiom. This is because, according to IIT, the phenomenological experience of an "irreducible whole" must be mirrored by physical irreducibility in the substrate that gives rise to consciousness. Because of this, \textit{physical feedback} is assumed to be a necessary (but not sufficient) condition for conscious experience, such that any system that lacks feedback has $\Phi=0$ by definition \cite{oizumi2014phenomenology}. Yet, results in automata theory suggest a different interpretation. In what follows, we point to the Krohn-Rhodes \cite{krohn1965algebraic,zeiger1967cascade} theorem and other feed-forward decomposition techniques \cite{doerig2019}, which prove that there is nothing inherently special about feedback from a functional perspective, aside from the fact that it often allows for a more efficient representation. This leads to the controversial conclusion that the behavior of any system with feedback and $\Phi>0$ can be perfectly emulated by a feed-forward philosophical zombie with $\Phi=0$. In response, IIT claims such systems lack consciousness because they are incapable of generating an integrated experience, but this claim rests solely on the assumption that physical feedback is the correct interpretation of what it means to have an integrated subjective experience.

To test this assumption, we demonstrate the existence of a fundamentally new type of feed-forward philosophical zombie, namely, one that is \textit{isomorphic} to its conscious counterpart in its state-transitions. The isomorphism guarantees that the feed-forward system with $\Phi=0$ not only emulates the input-output behavior of its conscious counterpart but does so \textit{without increasing the size of the system}. Thus, the \textit{internal} states of the two systems are in one-to-one correspondence, which allows us to isolate mathematical changes in $\Phi$ without introducing a qualitative difference in efficiency, autonomy, or behavior across systems. Indeed, we show the only mathematical difference between the "conscious" system with $\Phi>0$ and "unconscious" system with $\Phi=0$ is a permutation of the binary labels used to instantiate internal states. This implies $\Phi$ depends on the specific internal representation of a computation rather than the computation itself.  Our formalism translates into a proposed mathematical criterion that any measure of consciousness must be invariant under isomorphisms in the state transition diagram. Enforcement of this criterion serves as a necessary, but not sufficient, formal condition for any theory of consciousness to be free from philosophical zombies and the epistemological problems they pose.

\section{Methods} \label{methods}

\subsection{Finite-State Automata}
Finite-state automata are abstract computing devices, or "machines", designed to model a discrete system as it transitions between states. Automata theory was invented to address biological and psychological problems \cite{arbib1968algebraic, shannon2016automata} and it remains an extremely intuitive choice for modeling neuronal systems. This is because one can define an automaton in terms of how specific abstract inputs lead to changes within a system. Namely, if we have a set of potential inputs \(\Sigma\) and a set of internal states \(Q\), we define an automaton \(A\) in terms of the tuple \(A = (\Sigma,Q,\delta,q_0)\) where \(\delta:\Sigma \times Q \rightarrow Q\) is a map from the current state and input symbol to the next state, and \(q_0 \in Q\) is the starting state of the system. To simplify notation, we write \(\delta(s,q) = q'\) to denote the transition from \(q\) to \(q'\) upon receiving the input symbol \(s \in \Sigma\).

\begin{figure}[ht]

    \begin{subfigure}[b]{0.30\textwidth}
    \includegraphics[width=0.9\linewidth]{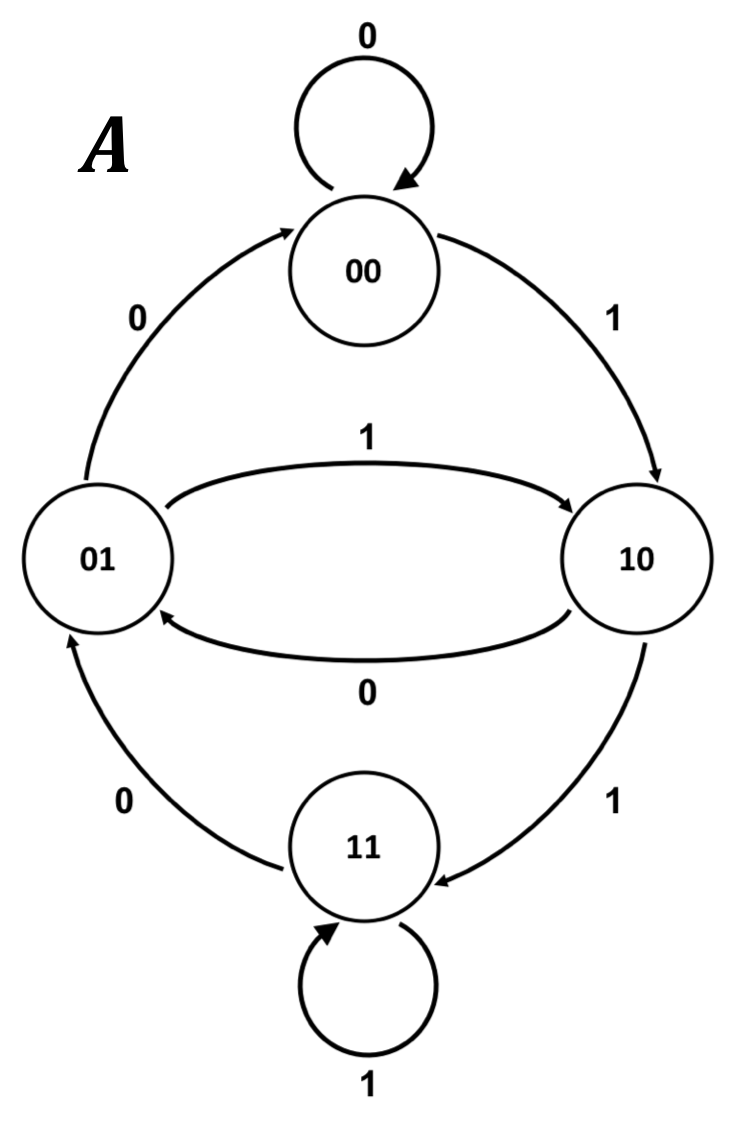} 
    \caption{}
    \label{fig:rightshift_a}
    \end{subfigure}
    \begin{subfigure}[b]{0.23\textwidth}
    \includegraphics[width=0.9\linewidth]{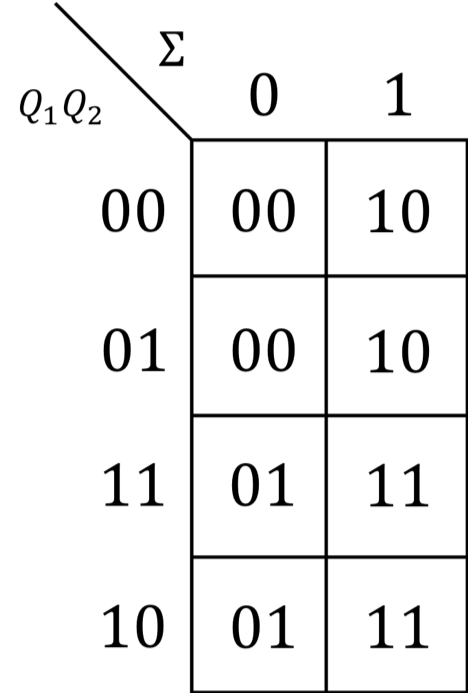}
    \caption{}
    \label{fig:rightshift_b}
    \end{subfigure}
    \begin{subfigure}[b]{0.47\textwidth}
    \includegraphics[width=0.9\linewidth]{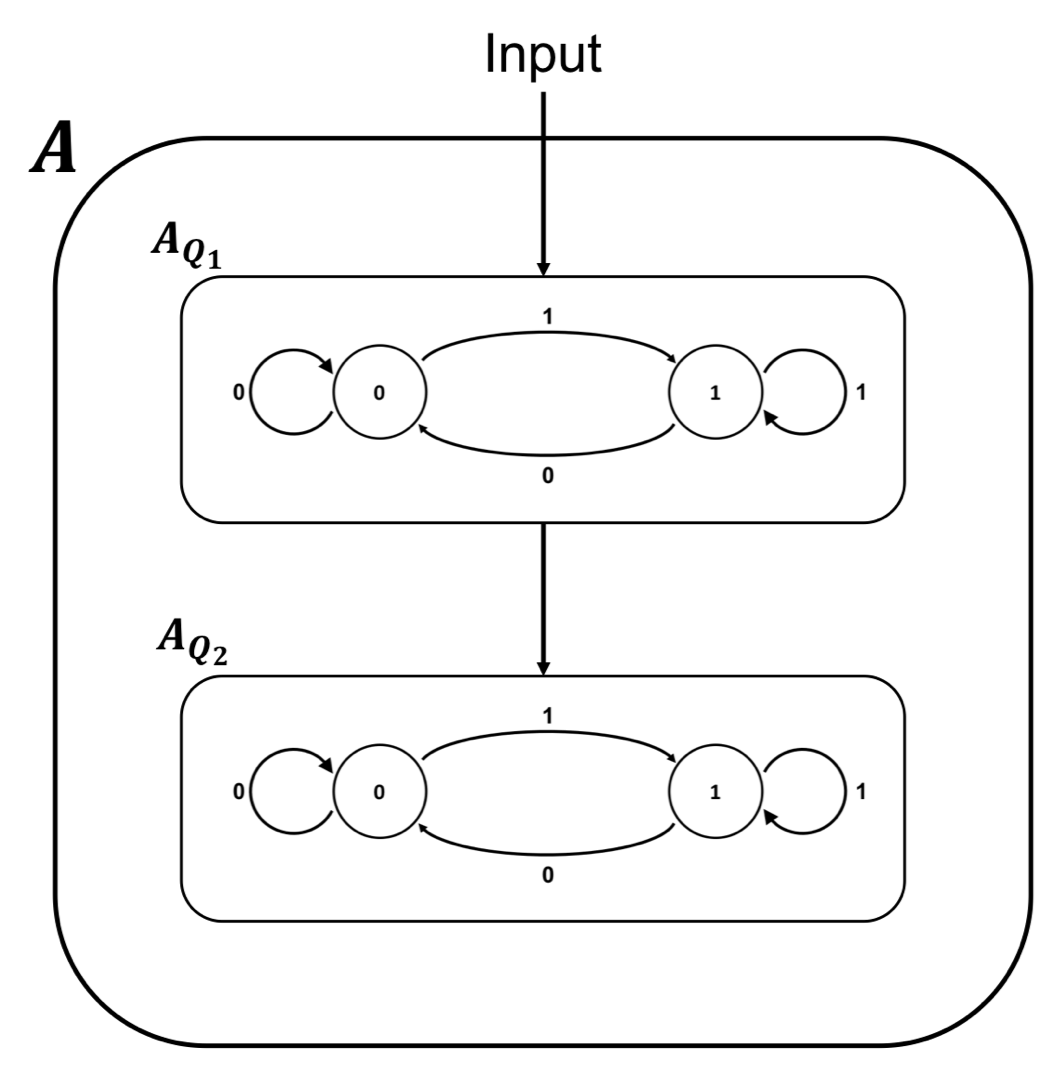}
    \caption{}
    \label{fig:rightshift_c}
    \end{subfigure}
    
    \caption{The "right-shift automaton" $A$ in terms of its state-transition diagram (\ref{fig:rightshift_a}), transition function $\delta$ (\ref{fig:rightshift_b}), and logical architecture (\ref{fig:rightshift_c}).}
    \label{fig:rightshift_automata}
\end{figure}

For example, consider the "right-shift automaton" $A$ shown in Figure \ref{fig:rightshift_automata}. This automaton is designed to model a system with a two bit internal register that processes new elements from the input alphabet \(\Sigma = \{0,1\}\) by shifting the bits in the register to the right and appending the new element on the left \cite{Dedeo2012}. The global state of the machine is the combined state of the left and right register, so \( Q = \{00,01,10,11\}\) and the transition function $\delta$ specifies how this global state changes in response to each input, as shown in Figure \ref{fig:rightshift_b}. 

In addition to the global state transitions, each individual bit in the register of the right-shift automaton is itself an automaton. In other words, the global functionality of the system is nothing more than the combined output from a system of interconnected automata, each specifying the state of a single component or "coordinate" of the system. Specifically, the right-shift automaton is comprised of an automaton $A_{Q1}$ responsible for the left bit of register and an automaton $A_{Q2}$ responsible for the right bit of the register. By definition, $A_{Q_1}$ copies the input from the environment and $A_{Q_2}$ copies the state of $A_{Q1}$. Thus, $\Sigma_{Q_1} = \{0,1\}$ and $\Sigma_{Q_2} = Q_1 = \{0,1\}$ and the transition functions for the coordinates are $\delta_{Q_1}=\delta_{Q_2} = \{\delta(0,0)=0; \delta(0,1)=0; \delta(1,0)=1; \delta(1,1)=1\}$. This fine-grained view of the right-shift automaton specifies its \textit{logical architecture} and is shown in Figure \ref{fig:rightshift_c}. The logical architecture of the system is the "circuitry" that underlies its behavior and, as such, is often specified explicitly in terms of logic gates, with the implicit understanding that each logic gate is a component automaton. 

It is important to note that not all automata require multiple input symbols and it is common to find examples of automata with a single-letter input alphabet. In fact, any deterministic state-transition diagram can be represented in this way, with a single input letter signaling the passage of time. In this case, the states of the automaton are the states of the system, the input alphabet is the passage of time, and the transition function $\delta$ is given by the transition probability matrix (TPM) for the system. Because $\Phi$ is a mathematical measure that takes a TPM as input, this specialized case provides a concrete link between IIT and automata theory. Non-deterministic TPMs can also be described in terms of finite-state automata \cite{maler1995decomposition, Dedeo2012} but, for our purposes, this generalization is not necessary.


\subsection{Cascade Decomposition} \label{Connection to IIT}

The idea of decomposability is central to both IIT and automata theory. As \citet{tegmark2016} points out, mathematical measures of integrated information, including $\Phi$, quantify the inability to decompose a transition probability matrix \(M\) into two independent processes \(M_A\) and \(M_B\). Given a distribution over initial states \(p\), if we approximate \(M\) by the tensor factorization \(\hat{M} \approx M_A \otimes M_B\), then \(\Phi\), in general, quantifies an information theoretic distance \(D\) between the regular dynamics \(Mp\) and the dynamics under the partitioned approximation \(\hat{M}p\) (i.e. $\Phi = D(Mp||\hat{M}p)$). In the latest version of IIT \cite{oizumi2014phenomenology}, only \textit{unidirectional} partitions are implemented (information can flow in one direction across the partition) which mathematically enforces the assumption that feedback is a necessary condition for consciousness.

Decomposition in automata theory, on the other hand, has historically been an engineering problem. The goal is to decompose an automaton \(A\) into an automaton \(A'\) which is made of simpler physical components than \(A\) and maps \emph{homomorphically} onto \(A\). Here, we define a homomorphism \(h\) as a map from the states, stimuli, and transitions of \(A'\) onto the states, stimuli, and transitions of \(A\) such that for every state and stimulus in \(A'\) the results obtained by the following two methods are equivalent \cite{arbib1968algebraic}:
\begin{enumerate}
    \item Use the stimulus of \(A'\) to update the state of \(A'\) then map the resulting state onto \(A\).
    \item Map the stimulus of \(A'\)and the state of \(A'\) to the corresponding stimulus/state in \(A\) then update the state of \(A\) using the stimulus of \(A\).
\end{enumerate}

\begin{figure}[ht]
\vspace{1ex}
\centering
\includegraphics[width=0.7\columnwidth]{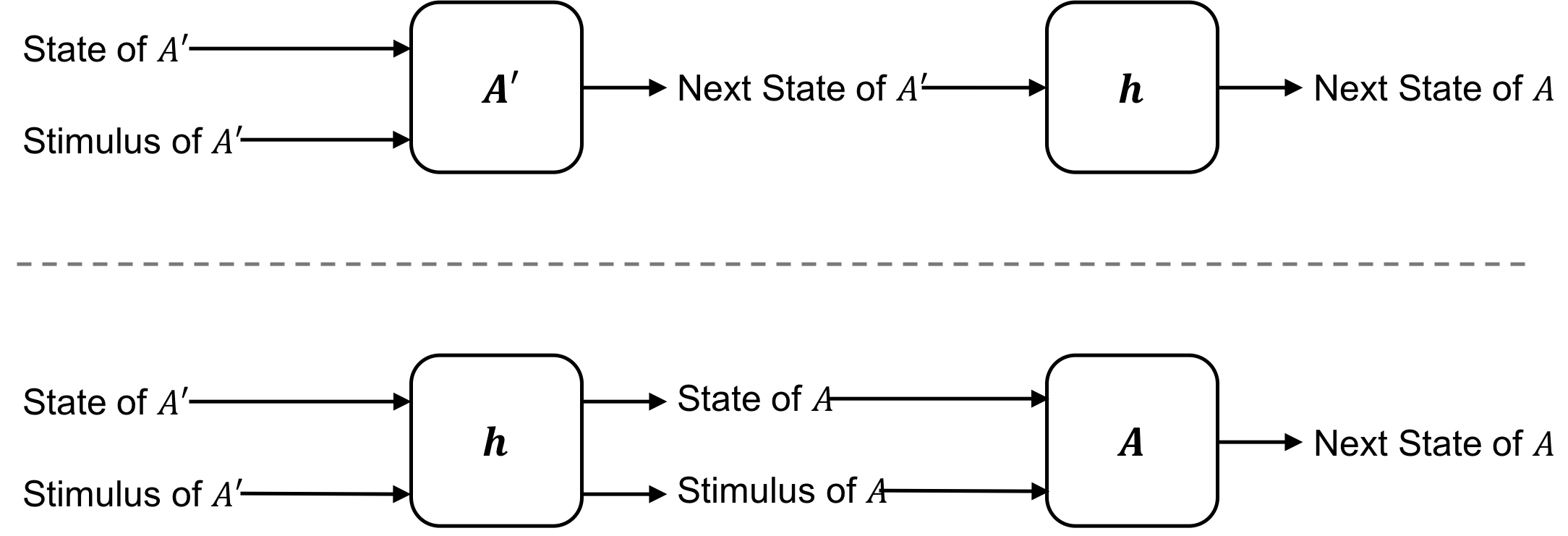}
\caption{For the map $h$ to be a homomorphism from $A'$ onto $A$, updating the dynamics then applying $h$ (top) must yield the same state of $A$ as applying $h$ then updating the dynamics (bottom).}
\label{fig:homomorphism}
\vspace{1ex}
\end{figure}

In other words, the map $h$ is a homomorphism if it \textit{commutes} with the dynamics of the system. The two operations (listed above) that must commute are shown schematically in Figure \ref{fig:homomorphism}. If the homomorphism $h$ is bijective then it is also an \emph{isomorphism} and the two automata necessarily have the the same number of states. 

From an engineering perspective, homomorphic/isomorphic logical architectures are useful because they allow flexibility when choosing a logical architecture to implement a given computation. Mathematically, the difference between homomorphic automata is the internal labeling scheme used to encode the states/stimuli of the global finite-state machine, which specifies the behavior of the system. Thus, the homomorphism $h$ is a dictionary that translates between different representations of the same computation. Just as the same sentence can be spoken in different languages, the same computation can be instantiated using different encodings. Under this view, what gives a computational state meaning is not its binary representation (label) but rather its causal relationship with other states/stimuli, which is what the homomorphism preserves. 


\begin{figure}[ht]
\centering
\includegraphics[width=0.7\columnwidth]{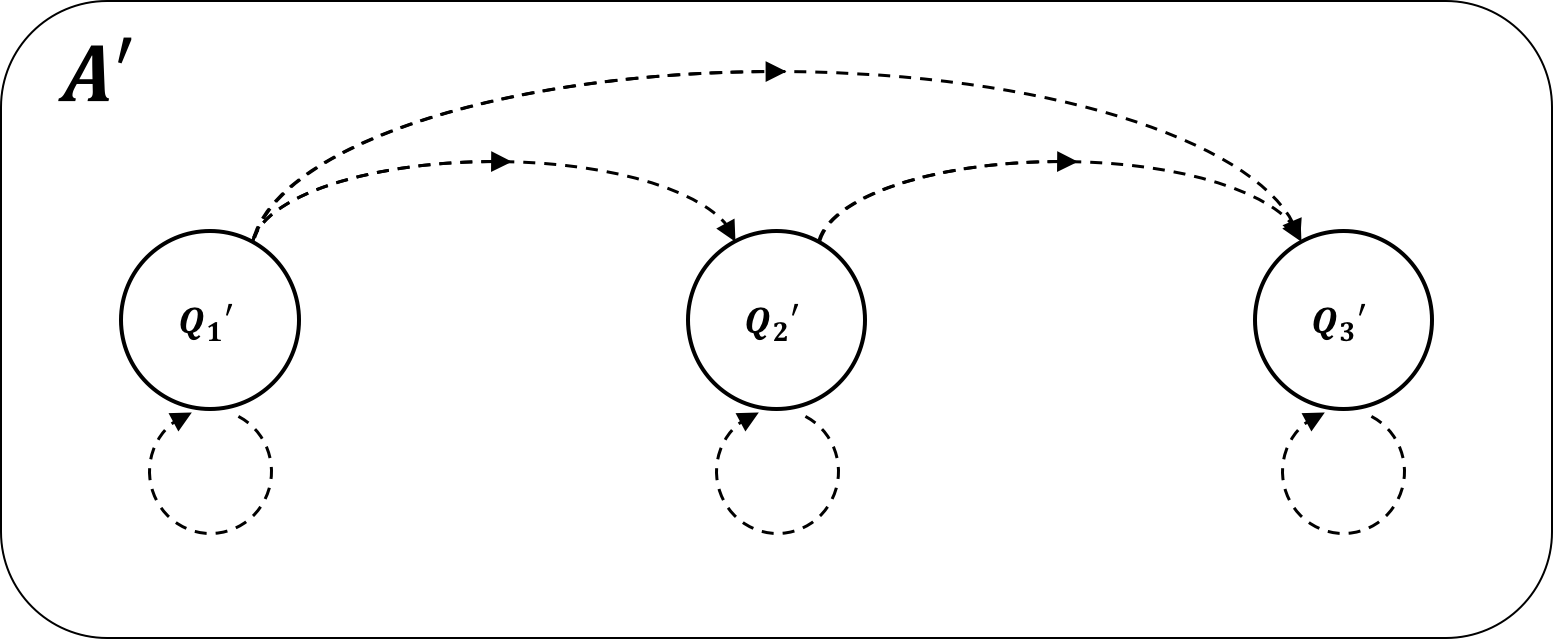}
\caption{An example of a fully connected three component system in cascade form. Any subset of the connections drawn above meets the criteria for cascade form because information flows unidirectionally.}
\label{fig:cascade_form}
\end{figure}

Because we are interested in isolating the role of feedback, the specific type of decomposition we seek is a feed-forward or \textit{cascade decomposition} of the logical architecture of a given system. Cascade decomposition takes the automaton $A$ and decomposes it into a homomorphic automaton $A'$ comprised of several elementary automata "cascaded together". By this, what is meant is that the output from one component serves as the input to another such that the flow of information in the system is strictly unidirectional (Figure \ref{fig:cascade_form}). The resulting logical architecture is said to be in "cascade" or "hierarchical" form and is functionally identical to the original system (i.e. it realizes the same global finite-state machine).

At this point, the connection between IIT and cascade decomposition is readily apparent: if an automaton with feedback allows a homomorphic cascade decomposition, then the behavior of the resulting system is indistinguishable from the original but utilizes only feed-forward connections. Therefore, there exists a unidirectional partition of the system that leaves the dynamics of the new system (i.e. the transition probability matrix) unchanged such that $\Phi=0$ for all states. 

In the language of \citet{oizumi2014phenomenology}, we can prove this by letting $C_{\rightarrow}$ be the constellation that is generated as a result of any unidirectional partition and $C$ be the original constellation. Because $C_{\rightarrow}$ has no effect on the TPM, we are guaranteed that $C_{\rightarrow}=C$ and $\Phi^{MIP}=D(C|C_\rightarrow)=0$. We can repeat this process for every possible subsystem within a given system and, since the flow of information is always unidirectional, $\Phi^{MIP}=0$ for all subsets so $\Phi^{Max}=0$. Thus, $\Phi=0$ for all states and subsystems of a cascade automaton.

Pertinently, the Krohn-Rhodes theorem proves that every automaton can be decomposed into cascade form \cite{krohn1965algebraic, zeiger1967cascade}, which implies \textit{every system for which we can measure non-zero $\Phi$ allows a feed-forward decomposition with $\Phi=0$}. These feed-forward systems are "philosophical zombies" in the sense that they lack subjective experience according to IIT (e.g. $\Phi=0$), but they nonetheless perfectly emulate the behavior of conscious systems. Yet, the Krohn-Rhodes theorem does not tell us \textit{how} to construct such systems. Furthermore, the map between systems is only guaranteed to be homomorphic (many-to-one) which allows for the possibility that $\Phi$ is picking up on other properties (e.g. such as the efficiency and/or autonomy of the computation) in addition to the presence or absence of feedback \cite{oizumi2014phenomenology}. 

To isolate what $\Phi$ is measuring, we must go one step further and insist that the decomposition is \textit{isomorphic} (one-to-one) such that the original and zombie systems can be considered to perform the same computation (same global state-transition topology) under the same resource constraints. In this case, the feed-forward system has the \textit{exact same number of states} as its counterpart with feedback. Provided the latter has $\Phi>0$, this implies $\Phi$ is not a measure of the efficiency of a given computation, as both systems require the same amount of memory. This is not to say that feedback and $\Phi$ do not \textit{correlate} with efficiency because, in general, they do \cite{albantakis2014}. For certain computations, however, the presence of feedback is not associated with increased efficiency but only increased interdependence among elements.

It is these specific corner cases that are most beneficial if one wants to assess the validity of the theory, as they allow one to understand whether or not feedback is important in absence of the benefits typically associated with its presence. In other words, IIT's translation of the integration axiom is that \textit{feedback} is a minimal criterion for the subjective experience of a unified whole; yet, $\Phi$ is described as quantifying "the amount of information generated by a complex of elements, above and beyond the information generated by its parts" \cite{tononi2008}, which seems to imply feedback enables something "extra" feed-forward systems cannot reproduce. An isomorphic feed-forward decomposition allows us to carefully track the mathematical changes that destroy this additional information, in a way that lets us preserve the efficiency and functionality of the original system. This, in turn, provides the clearest possible case to test whether or not this additional information is likely to correspond to a phenomenological difference between systems.

\subsection{Feed-forward Isomorphisms via Preserved Partitions} \label{Preserved Partitions}

The special type of computation that allows an isomorphic feed-forward decomposition is one in which the global state-transition diagram is amenable to decomposition via a nested sequence of preserved partitions. A preserved partition is a way of partitioning the state space of a system into blocks of states (macrostates) that transition together. Namely, a partition $P$ is preserved if it breaks the state space $S$ into a set of blocks $\{B_1, B_2, ... , B_N\}$ such that every state within each block transitions to a state within the same block \cite{hartmanis1966algebraic, arbib1968algebraic}. If we denote the state-transition function $f:S \rightarrow S$, then a block $B_i$  is preserved when:
$$
\exists j \in \{1,2,...,N\} \textrm{ such that } f(x) \in B_j \forall x \in B_i
$$
In other words, for $B_i$ to be preserved, $\forall x$ in $B_i$ $x$ must transition to some state in a single block $B_j$ ($i=j$ is allowed). Conversely, $B_i$ is \textit{not} preserved if there exist two or more states in $B_i$ that transition to different blocks (i.e. $\exists$ $x_1,x_2\in B_i$ such that $f(x_1)=B_j$ and $f(x_2)=B_k$ with $j \neq k$ ). In order for the entire partition $P_i$ to be preserved, each block within the partition must be preserved.

For an isomorphic cascade decomposition to exist, we must be able to iteratively construct a hierarchy or "nested sequence" of preserved partitions such that each partition $P_i$ evenly splits the partition $P_{i-1}$ above it in half, leading to a more and more refined description of the system.  For a system with $2^n$ states where $n$ is the number of binary components in the original system, this implies that we need to find exactly $n$ nested preserved partitions, each of which then maps onto a unique component of the cascade automaton, as demonstrated in Section \ref{example_section}. 

If one cannot find a preserved partition made of disjoint blocks or the blocks of a given partition do not evenly split the blocks of the partition above it in half, then the system in question does not allow an isomorphic feed-forward decomposition. It will, however, still allow a \textit{homomorphic} feed-forward decomposition based on a nested sequence of preserved \textit{covers}, which forms the basis of standard Krohn-Rhodes decomposition techniques \cite{arbib1968algebraic,egri2008hierarchical,egri2015computational}. Unfortunately, we do not know of a way to tell a priori whether or not a given computation will ultimately allow an isomorphic feed-forward decomposition, although a high degree of symmetry in the global state-transition diagram is certainly a requirement.

\subsubsection{Example: \texttt{AND}/\texttt{OR} $\cong$ \texttt{COPY}/\texttt{OR}} \label{example_section}

\begin{figure}[ht]
    \centering
    \begin{subfigure}[b]{0.4\textwidth}
    \centering
    \includegraphics[width=0.9\linewidth]{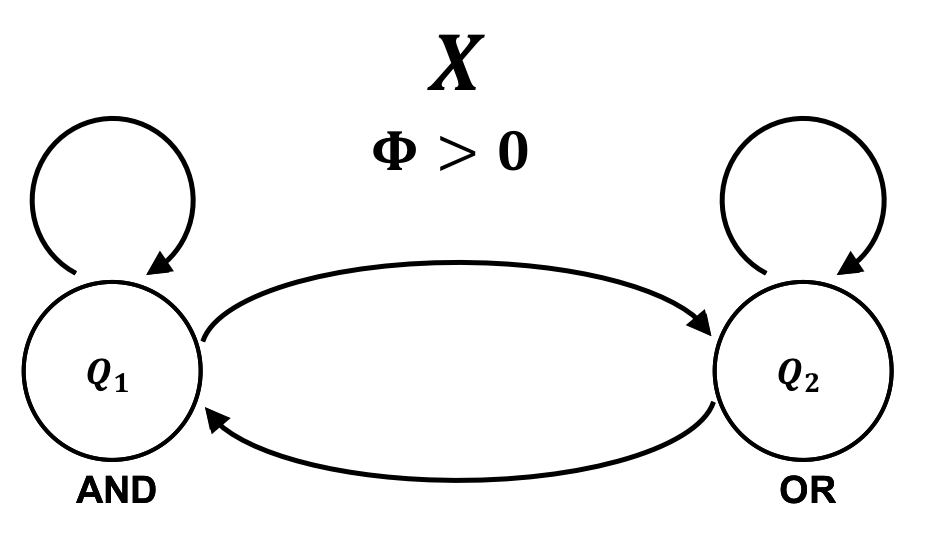}
    \caption{}
    \label{fig:X}
    \vspace{2ex}
    \includegraphics[width=0.9\linewidth]{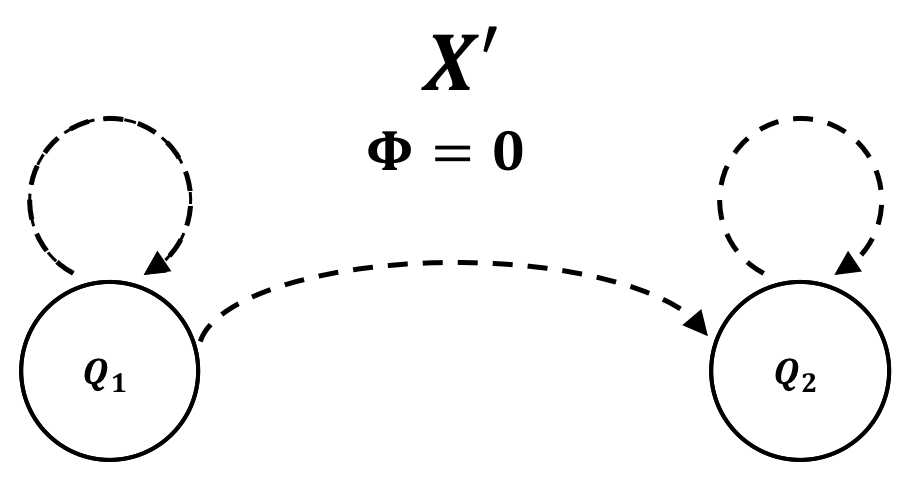}
    \caption{}
    \label{fig:X_prime}
    \end{subfigure}
    \begin{subfigure}[b]{0.4\textwidth}
    \centering
    \includegraphics[width=0.9\linewidth]{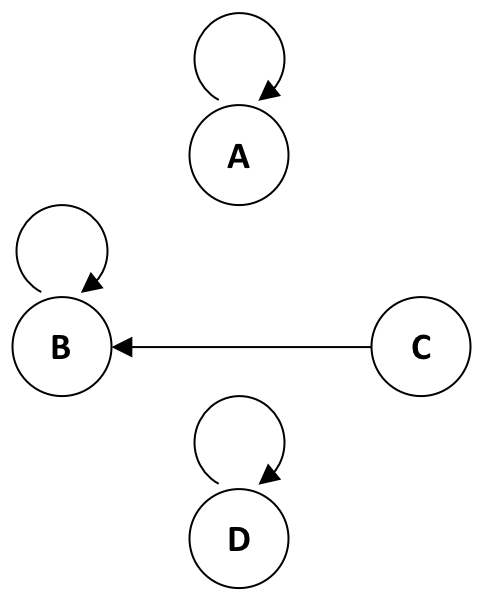}
    \caption{}
    \label{fig:X_topology}
    \end{subfigure}
    
    \caption{The goal of an isomorphic cascade decomposition is to decompose the integrated logical architecture of the system $X$ (\ref{fig:X}) so that it is in cascade form $X'$ (\ref{fig:X_prime}) without affecting the state-transition topology of the original system (\ref{fig:X_topology}).}
    \label{fig:rightshift}
\end{figure}

As an example, we will isomorphically decompose the feedback system $X$, comprised of an \texttt{AND} gate and an \texttt{OR} gate, shown in Figure \ref{fig:X}. As it stands, $X$ is not in cascade form because information flows bidirectionally between the components $Q_1$ and $Q_2$. While this feedback alone is insufficient to guarantee $\Phi>0$, one can readily check that $X$ does indeed have $\Phi>0$ for all possible states (e.g. \cite{mayner2018pyphi}). The global state-transition diagram for the system $X$ is shown in Figure \ref{fig:X_topology}. Note, we have purposefully left off the binary labels that $X$ uses to instantiate these computational states, as the goal is to relabel them in a way that results in a different (feed-forward) instantiation. In general, one typically starts from the computation and derives a single logical architecture but, here, we must start and end with fixed (isomorphic) logical architectures - passing through the underlying computation in between. The general form of the feed-forward logical architecture $X'$ that we seek is shown in Figure \ref{fig:X_prime}. 

\begin{figure}[ht]
    \centering
    
    \begin{subfigure}[b]{0.38\textwidth}
    \includegraphics[width=0.9\linewidth]{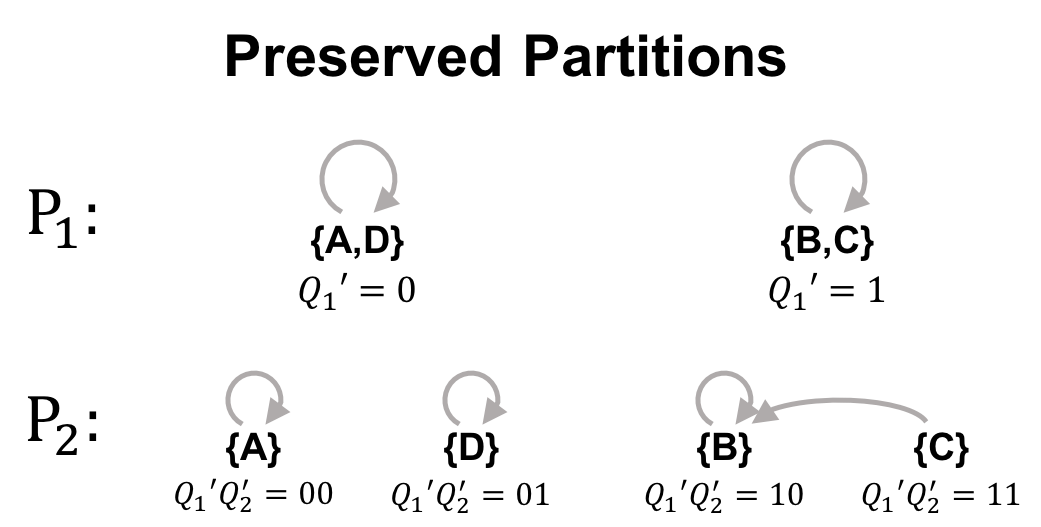}
    \caption{}
    \label{fig:X_partitions}
    \end{subfigure}
    \begin{subfigure}[b]{0.33\textwidth}
    \includegraphics[width=0.9\linewidth]{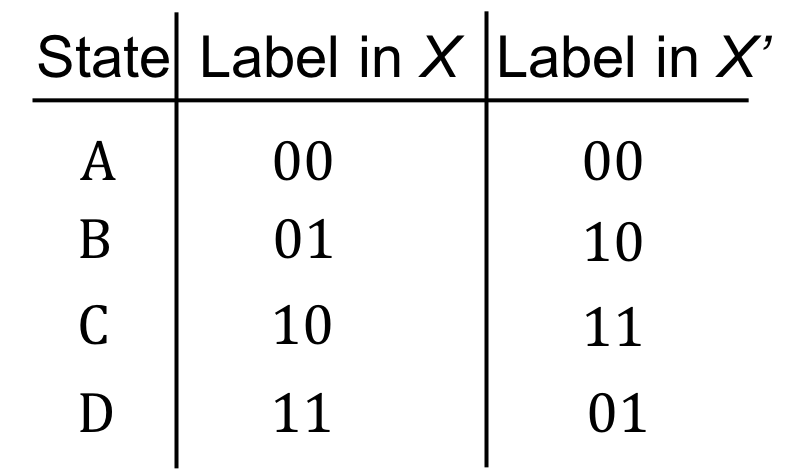}
    \caption{}
    \label{fig:X_isomorphism}
    \end{subfigure}
    \begin{subfigure}[b]{0.27\textwidth}
    \includegraphics[width=0.9\linewidth]{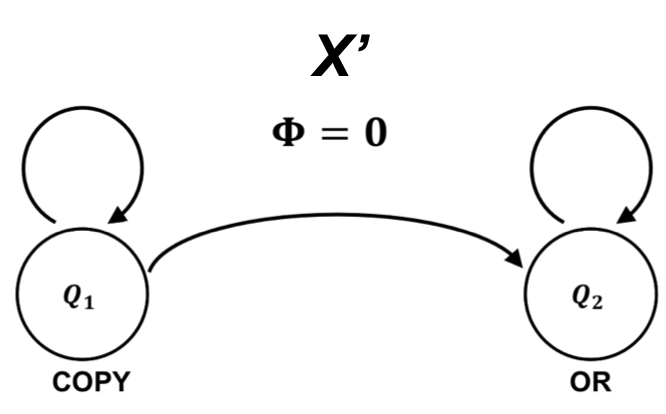}
    \caption{}
    \label{fig:Xnew_prime}
    \end{subfigure}
    
    \caption{The nested sequence of preserved partitions in \ref{fig:X_partitions} yields the isomorphism \ref{fig:X_isomorphism} between $X$ and $X'$ which can be translated into the strictly feed-forward  logical architecture with $\Phi=0$ shown in \ref{fig:Xnew_prime}.}
    \label{fig:X_decomp}
\end{figure}

Given the global state-transition diagram shown in Figure \ref{fig:X_topology}, we let our first preserved partition be $P_1=\{B_0,B_1\}$ with $B_0=\{A,D\}$ and $B_1=\{B,C\}$. It is easy to check that this partition is preserved, as one can verify that every element in $B_0$ transitions to an element in $B_0$ and every element in $B_1$ transitions to an element in $B_1$ (shown topologically in Figure \ref{fig:X_partitions}). We then assign all the states in $B_0$ a first coordinate value of $0$ and all the states in $B_1$ a first coordinate value of $1$, which guarantees the state of the first coordinate is independent of later coordinates. If the value of the first coordinate is $0$ it will remain $0$ and if the value of the first coordinate is $1$ it will remain one, because states within a given block transition together. Because $0$ goes to $0$ and $1$ goes to $1$, the logic element (component automaton) representing the first coordinate $Q_1'$ is a \texttt{COPY} gate receiving its previous state as input.

The second preserved partition $P_2$ must evenly split each block within $P_1$ in half. Letting $P_2 = \{\{B_{00},B_{01}\},\{B_{10},B_{11}\}\}$ we have $B_{00}=\{A\}$, $B_{01}=\{B\}$, $B_{10}=\{C\}$, and $B_{11}=\{D\}$. At this stage, it is trivial to verify that the partition is preserved because each block is comprised of only one state which is guaranteed to transition to a single block. As with $Q_1'$, the logic gate for the second coordinate ($Q_2'$) is specified by the way the labeled blocks of $P_2$ transition. Namely, we have $B_{00} \rightarrow B_{00} $, $B_{01} \rightarrow B_{01}$, $B_{10} \rightarrow B_{01}$, and $B_{11} \rightarrow B_{11}$. Note, the transition function $\delta_{Q2}$ is completely deterministic given input from the first two coordinates (as required) and is given by $\delta_{Q2}=\{ 00\rightarrow 0; 01 \rightarrow 1; 10 \rightarrow 1; 11 \rightarrow 1\}$. This implies $Q_2'$ is an \texttt{OR} gate receiving input from both $Q_1'$ and $Q_2'$.

At this point, the isomorphic cascade decomposition is complete. We have constructed an automaton for $Q_1'$ that takes input from only itself and an automaton for $Q_2'$ that takes input only from itself and earlier coordinates (i.e. $Q_1'$ and $Q_2'$). The mapping between the states of $X$ and the states of $X'$, shown in Figure \ref{fig:X_isomorphism}, is specified by identifying the binary labels (internal representations) each system uses to instantiate the abstract computational states $A,B,C,D$ of the global state-transition diagram. Because $X$ and $X'$ operate on the same support (the same four binary states) the fact that they are isomorphic implies the difference between representations is nothing more than a permutation of the labels used to instantiate the computation. By choosing a specific labelling scheme based on isomorphic cascade decomposition, we can induce a logical architecture that is guaranteed to be feedback free and has $\Phi=0$. In this way we have "unfolded" the feedback present in $X$ without affecting the size/efficiency of the system.

\section{Results/Discussion}

\begin{figure}[ht]
    \centering
    
    \begin{subfigure}[t]{0.35\textwidth}
    \includegraphics[width=0.9\linewidth]{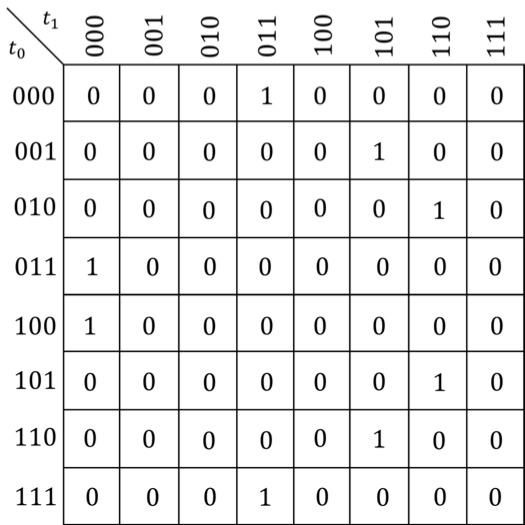}
    \caption{}
    \label{fig:Y_TPM}
    \end{subfigure}
    \begin{subfigure}[t]{0.35\textwidth}
    \includegraphics[width=0.9\linewidth]{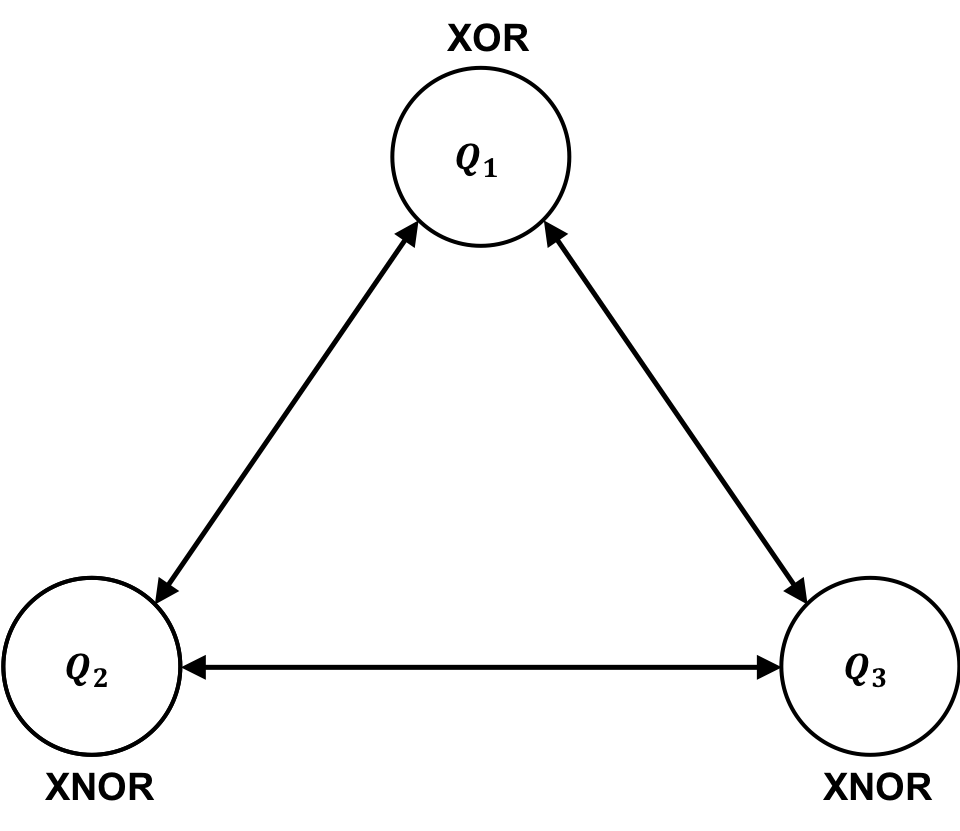}
    \caption{}
    \label{fig:Y_diagram}
    \end{subfigure}
    \begin{subfigure}[t]{0.2\textwidth}
    \includegraphics[width=0.9\linewidth]{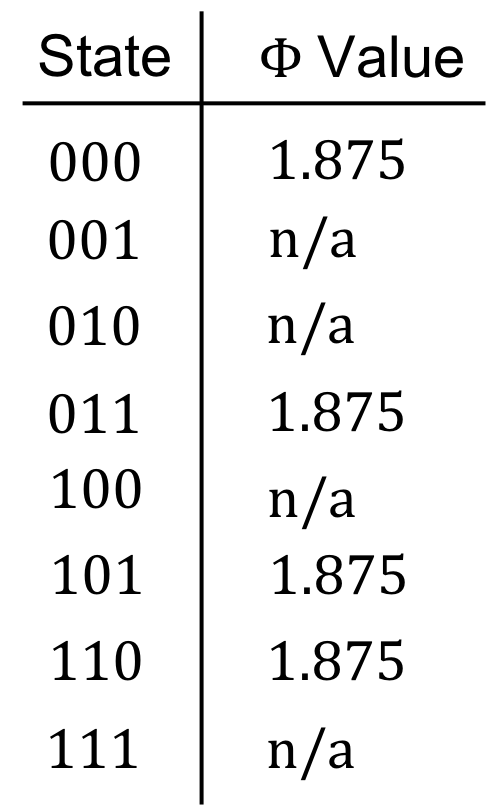}
    \caption{}
    \label{fig:Y_transitions}
    \end{subfigure}
    
    \caption{The transition probability matrix (\ref{fig:Y_TPM}), logical architecture (\ref{fig:Y_diagram}), and all available $\Phi$ values (\ref{fig:Y_transitions}) for the example system $Y$ (n/a implies $\Phi$ is not defined for a given state because it is unreachable).}
    \label{fig:Y_system}
\end{figure}

\begin{figure}[!ht]
    \centering
    
    \begin{subfigure}[t]{0.4\textwidth}
    \centering
    \includegraphics[width=0.7\linewidth]{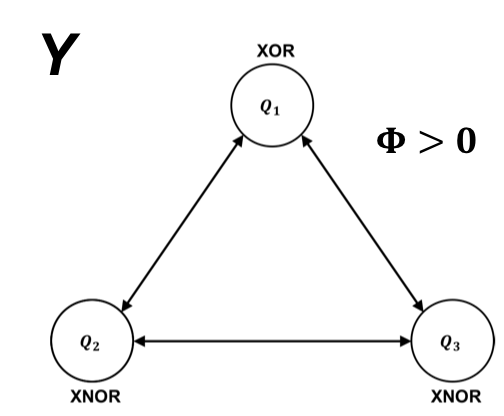}
    \caption{}
    \label{fig:Y_A}
    \vspace{2ex}
    \includegraphics[width=0.35\linewidth]{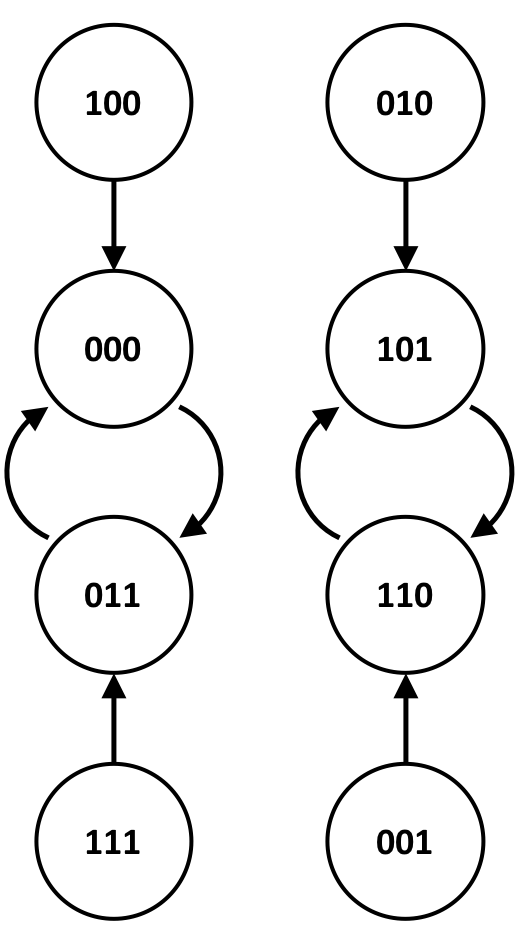}
    \caption{}
    \label{fig:Y_B}
    \end{subfigure}
    \begin{subfigure}[t]{0.4\textwidth}
    \centering
    \includegraphics[width=0.9\linewidth]{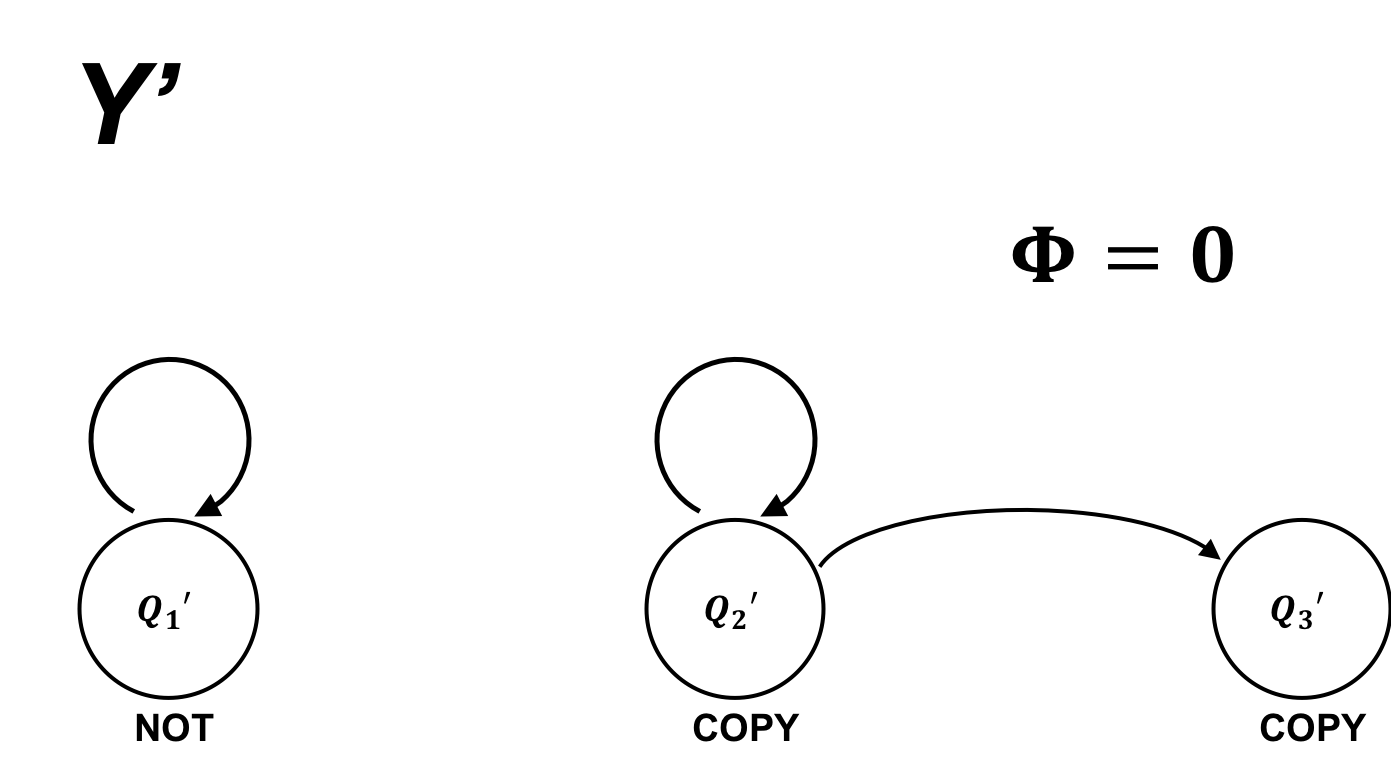}
    \caption{}
    \label{fig:Y_C}
    \vspace{2ex}
    \includegraphics[width=0.35\linewidth]{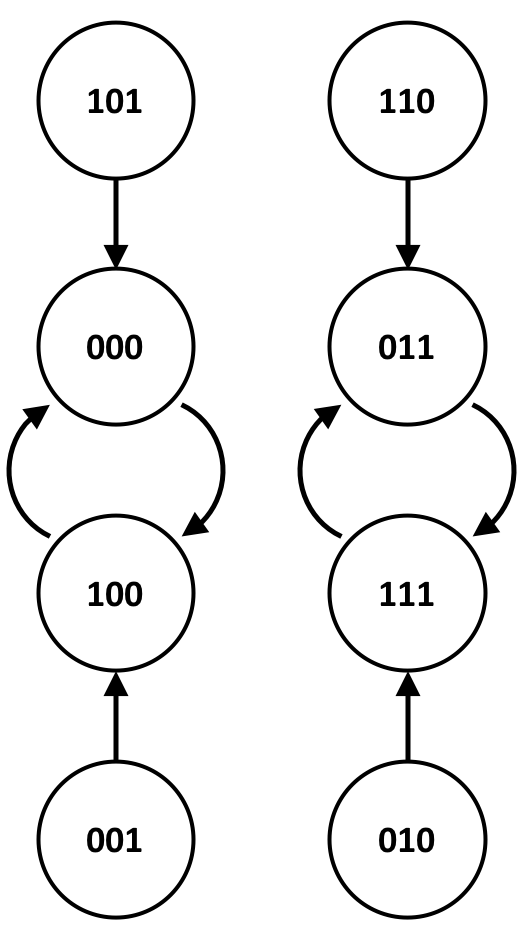}
    \caption{}
    \label{fig:Y_D}
    \end{subfigure}
    
    \caption{Side-by-side comparison of the feedback system $Y$ with $\Phi>0$ (\ref{fig:Y_A}) and its isomorphic feed-forward counterpart $Y'$ with $\Phi=0$ (\ref{fig:Y_C}). The global state-transition diagrams (\ref{fig:Y_B} and \ref{fig:Y_D}, respectively) differ only by a permutation of labels.}
    \label{fig:Y_final}
\end{figure}

We are now prepared to demonstrate the existence of isomorphic feed-forward philosophical zombies in systems similar to those found in \citet{oizumi2014phenomenology}. To do so, we will decompose the integrated system $Y$ shown in Figure \ref{fig:Y_system} into an isomorphic feed-forward philosophical zombie $Y'$ of the form shown in Figure \ref{fig:cascade_form}. The system $Y$, comprised of two \texttt{XNOR} gates and one \texttt{XOR} gate, clearly contains feedback between components and has $\Phi>0$ for all states for which $\Phi$ can be calculated (Figure \ref{fig:Y_C}). As in Section \ref{Preserved Partitions}, the goal of the decomposition is an isomorphic relabeling of the finite-state machine representing the global behavior of the system, such that the induced logical architecture is strictly feed-forward.

\begin{figure}[ht]
    \centering
    \includegraphics[width=0.9\linewidth]{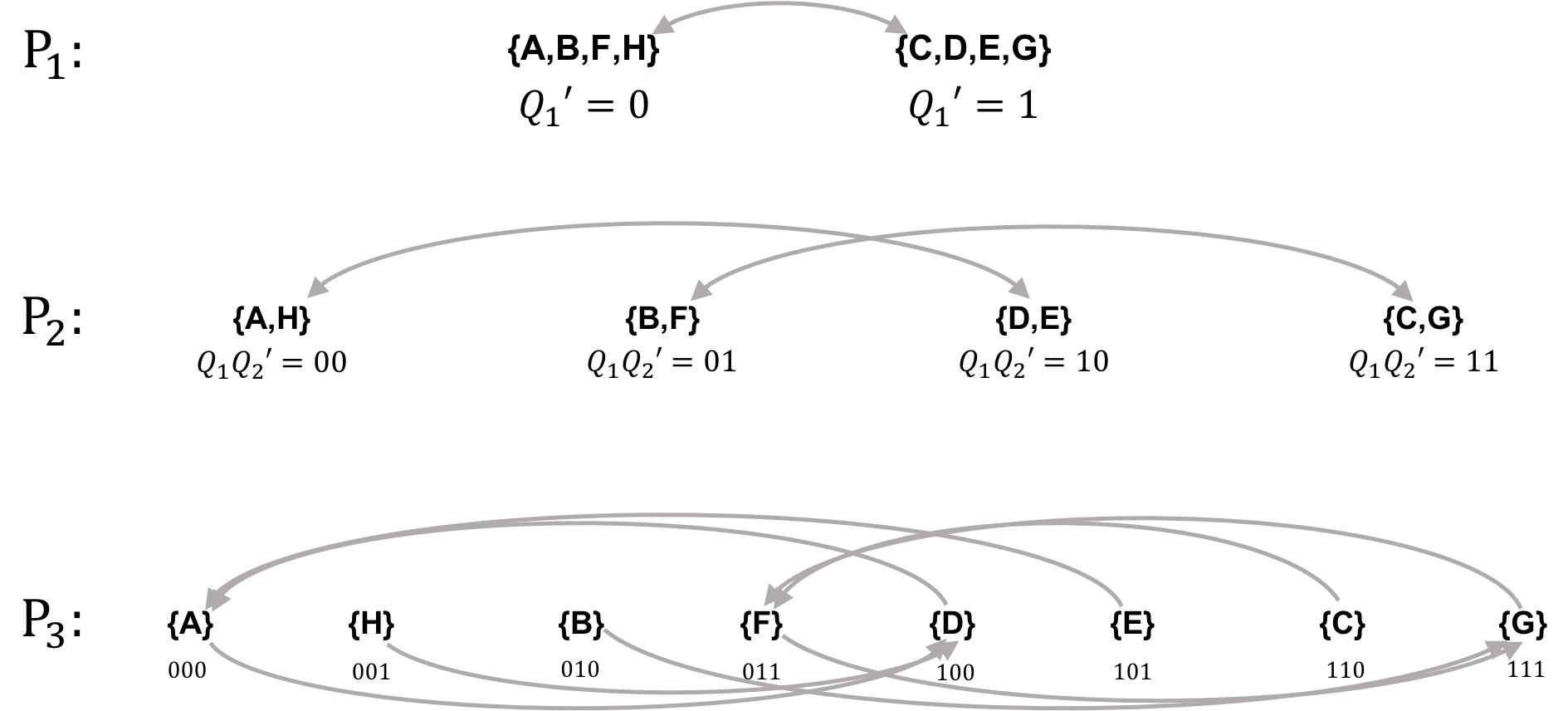}
    \caption{Nested sequence of preserved partitions used to decompose $Y$ into cascade form.}
    \label{fig:Y_partitions}
\end{figure}

We first evenly partition the state space of $Y$ into two blocks $B_0 = \{A,C,G,H\}$ and $B_1 = \{B,D,E,F\}$. Under this partition, $B_0$ transitions to $B_1$ and $B_1$ transitions to $B_0$, which implies the automaton representing the first coordinate in the new labeling scheme is a \texttt{NOT} gate. Note, this choice is not unique, as we could just as easily have chosen a different preserved partition such as $B_0 = \{A,D,E,H\}$ and $B_1 = \{B,C,F,G\}$, in which case the first coordinate would be a \texttt{COPY} gate; as long as the partition is preserved, the choice here is arbitrary and amounts to selecting one of several different feed-forward logical architectures - all in cascade form. For the second preserved partition, we let $P_2 = \{\{B_{00},B_{01}\},\{B_{10},B_{11}\}\}$ with $B_{00}=\{C,G\}$, $B_{01}=\{A,H\}$, $B_{10}=\{B,F\}$. and $B_{11}=\{D,E\}$. The transition function for the automaton representing the second coordinate, given by the movement of these blocks, is: $\delta_{Q2'}=\{ 00\rightarrow 0; 01 \rightarrow 1; 10 \rightarrow 0; 11 \rightarrow 1\}$, which is again a \texttt{COPY} gate receiving input from itself. The third and final partition $P_3$ assigns each state to its own unique block. As is always the case, this last partition is trivially preserved because individual states are guaranteed to transition to a single block. The transition function for this coordinate, read off the bottom row of Figure \ref{fig:Y_partitions}, is given by:
$$
\delta_{Q3'} = \{000 \rightarrow 0; 001 \rightarrow 0; 010 \rightarrow 1; 011 \rightarrow 1;  100 \rightarrow 0; 101 \rightarrow 0; 110 \rightarrow 1; 111 \rightarrow 1\}
$$
Using Karnuagh maps \cite{karnaugh1953map}, one can identify $\delta_{Q3}$ as a \texttt{COPY} gate receiving input from $Q_2'$. With the specification of the logic for the third coordinate, the cascade decomposition is complete and the new labeling scheme is shown in Figure \ref{fig:Y_partitions}. A side-by-side comparison of the original system $Y$ and the feed-forward system $Y'$ is shown in Figure \ref{fig:Y_final}. As required, the feed-forward system has $\Phi=0$ but executes the same sequence of state transitions as the original system, modulo a permutation of the labels used to instantiate the states of the global state-transition diagram. 
If the system with $\Phi>0$ is experiencing something more, it is something beyond the finite-state description of the system (the state-transition diagram) and, therefore, its presence or absence has no causal consequences to the structure of its internal state-transition map (the computation it performs).

\subsection{Discussion}
Behavior is most frequently described in terms of \textit{abstract} states/stimuli, which are not tied to a specific representation (binary or otherwise). Examples include descriptors of mental states, such as being asleep or awake, etc., these are representations of system states that must be defined either by an external observer or internally in the system performing the computation by its own logical implementation, but are not necessarily an intrinsic attribute of the computational states themselves (e.g. these states could be labeled with any binary assignment consistent with the state transition diagram of the computation). The analysis presented here is based on this premise, such that behavior is defined by the topology of the state-transition diagram, independent of a particular labeling scheme. And, indeed, it is this premise that enables Krohn-Rhodes decomposition to be useful from an engineering perspective, as one can swap between logical architectures without affecting the operation of a system in any way.

Phenomenologically, consciousness is often associated with the concept of ``top-down causation'', where `higher level' mental states exert causal control over lower level implementation \cite{ellis2016can}. The "additional information" provided by consciousness above and beyond non-conscious systems is considered to be functionally relevant by affecting how states transition to other states. It is in this sense behavior associated with consciousness in our formalism is most appropriately thought of as a computation having to do with the topology (causal architecture) of state-transitions, rather than the labels of the states or the specific logical architecture. We note this kind of  top-down causation can occur without the need to appeal to supervenience because in our framework the computation/function describes a functional equivalence class \cite{auletta2008top} of logical architectures that all implement the same causal relations among states, {\it  i.e.} it is an abstraction implemented in a particular logic. There is no additional ``room at the bottom'' for a particular logical architecture to exert more causal influence than another if they perform the same function. Any measure of consciousness that changes under the isomorphism we introduce here, such as $\Phi$, cannot therefore account for ``additional information'' in  this sense,  because of the existence of zombie systems with an identical structure to their state transitions. 

It is important to recognize there exists an alternative interpretation of behavior associated to consciousness, where one defines behavior not in terms of abstract computation, but in terms of the specific logical implementation, for example as IIT adopts. A clear example is the right-shift automaton, which is defined in terms of the relationship between components, resulting in a state-transition diagram with specific labels, because their exists strict constraints on the logical implementation of the behavior. However, this does not address, in our view, whether isomorphic systems ultimately experience a phenomenological difference as there is no way to test that assumption other than accepting it axiomatically. In particular, for the examples we consider here, there is only one input signal, meaning there are not multiple ways to encode input from the environment. Therefore there is no physical mechanism by which the environment can dictate a privileged internal representation. Instead the choice of internal representation is arbitrary with respect to the environment and depends only on the physical constraints of the architecture of the system performing the computation. For a system as complex as the human brain, there are presumably many possible logical architectures (network topologies) that can perform the same computation given the same input, differing only in how the states are internally represented ({\it e.g.} by how neurons are wired together). Why the ones that have evolved were selected in the first place is important for understanding why consciousness emerged in the universe. The foregoing suggests $\Phi$ is independent of functionality (computation), which implies there are no inherent evolutionary benefits to the presence or absence of $\Phi$ because it is not selectable as being distinctive to a particular computation an organism must perform for survival, only to how that computation is internally represented. 
Integration as a criterion separated from computation, is also itself problematic as it must be experienced by every component individually, yet, a single bit can make only one measurement (yes/no) and therefore components cannot sense where their information came from or where it is going.

This leads us to the central question of this manuscript,  which is what is experienced as the isomorphic system with $\Phi>0$ cycles through its internal states that is not experienced by the isomorphic system with $\Phi=0$? Since in our examples the environment is not dictating the representation of the input, and all state transitions are isomorphic, the representation and therefore the logic is arbitrary so long as a logical  architecture is selected with the proper input-output map under all circumstances.  
In light of this, our formalism suggests any mathematical measure of consciousness, phenomenologically motivated or otherwise, must be invariant under isomorphisms in the computation. This minimal criterion implies measurable differences in consciousness are always associated with measurable differences in the computational function (though the inverse need not be true), which is nothing more than a precise mathematical enforcement of the precedent set by Turing \cite{Turing1950}. From this perspective,  measures of consciousness should operate on the topology of the state-transition diagram, rather than the topology or logic of a particular physical implementation. That is, they should probe the computational capacity of the system without being biased by implementation - allowing identifying equivalence classes of physical systems that could have the same or similar conscious experience.


Our motivation in this work is to provide new roads to address the hard problem of consciousness by raising new questions. Our framework focuses attention on the fact that we currently lack a sufficiently formal understanding of the relationship between physical implementation and computation to truly address the hard problem. 
The logical architectures in Figures \ref{fig:Y_A} and \ref{fig:Y_C} are radically different, and yet, they perform the same computation. The fact that this computation allows a feed-forward decomposition is a consequence of redundancies that allow a compressed description in terms of a feed-forward logical architecture.  There are symmetries present in the computation that allow one to take advantage of shortcuts and reduce the computational load. This, in turn, shows up as a flexibility in the logical architecture that can generate the computation. In other words, the computation in question does not appear to require the maximum computational power of a three-bit logical architecture. For sufficiently complex eight-state computations, however, the full capacity of a three-bit architecture is required, as there is no redundancy to compress. Such systems cannot be generated without feedback, as the presence of feedback is accompanied by indispensable \textit{functional consequences}. Thus, the \textit{computation} is special because it cannot be efficiently represented without feedback - a relationship that can, in principle, be understood, but is only tangentially accounted for in current formalisms. It is up to the community to decide if the functionalist or phenomenological perspective will ultimately hold up, our goal in this work is simply to point out where the distinction between the two sets of ideas is very apparent and clear-cut mathematically, so that additional progress can be made.








\authorcontributions{Conceptualization, J.H. and S.W.; formal analysis, J.H.; funding acquisition, S.W.; investigation, J.H.; methodology, J.H.; project administration, S.W.; resources, S.W.; supervision, S.W.; visualization, J.H.; writing-original draft preparation, J.H. and S.W.; writing-review and editing, J.H. and S.W.}


\acknowledgments{The authors would like to thank Doug Moore for his assistance with the Krohn-Rhodes theorem and semigroup theory, as well as Dylan Gagler and the rest of the emergence@asu lab for thoughtful feed-back and discussions. SIW also acknowledges funding support from the Foundational Questions in Science Institute.}
\conflictsofinterest{The authors declare no conflict of interest. The funders had no role in the design of the study; in the collection, analyses, or interpretation of data; in the writing of the manuscript, or in the decision to publish the results.}





\reftitle{References}


\externalbibliography{yes}
\bibliography{references}

\begin{thebibliography}{-------}
\providecommand{\natexlab}[1]{#1}

\bibitem[Rees \em{et~al.}(2002)Rees, Kreiman, and Koch]{rees2002neural}
Rees, G.; Kreiman, G.; Koch, C.
\newblock Neural correlates of consciousness in humans.
\newblock {\em Nature Reviews Neuroscience} {\bf 2002}, {\em 3},~261.

\bibitem[Chalmers(1995)]{Chalmers1995}
Chalmers, D.J.
\newblock Facing Up to the Problem of Consciousness.
\newblock {\em Journal of Consciousness Studies} {\bf 1995}, {\em 2},~200--19.

\bibitem[Searle(1980)]{searle1980minds}
Searle, J.R.
\newblock Minds, brains, and programs.
\newblock {\em Behavioral and brain sciences} {\bf 1980}, {\em 3},~417--424.

\bibitem[Tononi(2008)]{tononi2008}
Tononi, G.
\newblock Consciousness as integrated information: a provisional manifesto.
\newblock {\em The Biological Bulletin} {\bf 2008}, {\em 215},~216--242.

\bibitem[Oizumi \em{et~al.}(2014)Oizumi, Albantakis, and
  Tononi]{oizumi2014phenomenology}
Oizumi, M.; Albantakis, L.; Tononi, G.
\newblock From the phenomenology to the mechanisms of consciousness: integrated
  information theory 3.0.
\newblock {\em PLoS computational biology} {\bf 2014}, {\em 10},~e1003588.

\bibitem[Tononi \em{et~al.}(2016)Tononi, Boly, Massimini, and
  Koch]{tononi2016integrated}
Tononi, G.; Boly, M.; Massimini, M.; Koch, C.
\newblock Integrated information theory: from consciousness to its physical
  substrate.
\newblock {\em Nature Reviews Neuroscience} {\bf 2016}, {\em 17},~450.

\bibitem[Shannon(1948)]{shannon1948}
Shannon, C.E.
\newblock A mathematical theory of communication.
\newblock {\em Bell system technical journal} {\bf 1948}, {\em 27},~379--423.

\bibitem[Tononi(2004)]{tononi2004information}
Tononi, G.
\newblock An information integration theory of consciousness.
\newblock {\em BMC neuroscience} {\bf 2004}, {\em 5},~42.

\bibitem[Balduzzi and Tononi(2008)]{balduzzi2008integrated}
Balduzzi, D.; Tononi, G.
\newblock Integrated information in discrete dynamical systems: motivation and
  theoretical framework.
\newblock {\em PLoS computational biology} {\bf 2008}, {\em 4},~e1000091.

\bibitem[Godfrey-Smith(2009)]{godfrey2009}
Godfrey-Smith, P.
\newblock {\em Theory and reality: An introduction to the philosophy of
  science}; University of Chicago Press,  2009.

\bibitem[Turing(1950)]{Turing1950}
Turing, A.
\newblock Computing Machinery and Intelligence.
\newblock {\em Mind} {\bf 1950}, {\em 59},~433--433.

\bibitem[Harnad(1995)]{harnad1995}
Harnad, S.
\newblock Why and how we are not zombies.
\newblock {\em Journal of Consciousness Studies} {\bf 1995}, {\em 1},~164--167.

\bibitem[Doerig \em{et~al.}(2019)Doerig, Schurger, Hess, and
  Herzog]{doerig2019}
Doerig, A.; Schurger, A.; Hess, K.; Herzog, M.H.
\newblock The unfolding argument: Why IIT and other causal structure theories
  cannot explain consciousness.
\newblock {\em Consciousness and Cognition} {\bf 2019}, {\em 72},~49 -- 59.
\newblock
  doi:{\changeurlcolor{black}\href{https://doi.org/https://doi.org/10.1016/j.concog.2019.04.002}{\detokenize{https://doi.org/10.1016/j.concog.2019.04.002}}}.

\bibitem[Krohn and Rhodes(1965)]{krohn1965algebraic}
Krohn, K.; Rhodes, J.
\newblock Algebraic theory of machines. I. Prime decomposition theorem for
  finite semigroups and machines.
\newblock {\em Transactions of the American Mathematical Society} {\bf 1965},
  {\em 116},~450--464.

\bibitem[Zeiger(1967)]{zeiger1967cascade}
Zeiger, H.P.
\newblock Cascade synthesis of finite-state machines.
\newblock {\em Information and Control} {\bf 1967}, {\em 10},~419--433.

\bibitem[Arbib \em{et~al.}(1968)Arbib, Krohn, and Rhodes]{arbib1968algebraic}
Arbib, M.; Krohn, K.; Rhodes, J.
\newblock {\em Algebraic theory of machines, languages, and semi-groups};
  Academic Press,  1968.

\bibitem[Shannon and McCarthy(2016)]{shannon2016automata}
Shannon, C.E.; McCarthy, J.
\newblock {\em Automata Studies.(AM-34)}; Vol.~34, Princeton University Press,
  2016.

\bibitem[DeDeo(2011)]{Dedeo2012}
DeDeo, S.
\newblock Effective Theories for Circuits and Automata.
\newblock {\em Chaos (Woodbury, N.Y.)} {\bf 2011}, {\em 21},~037106.
\newblock
  doi:{\changeurlcolor{black}\href{https://doi.org/10.1063/1.3640747}{\detokenize{10.1063/1.3640747}}}.

\bibitem[Maler(1995)]{maler1995decomposition}
Maler, O.
\newblock A decomposition theorem for probabilistic transition systems.
\newblock {\em Theoretical Computer Science} {\bf 1995}, {\em 145},~391--396.

\bibitem[Tegmark(2016)]{tegmark2016}
Tegmark, M.
\newblock Improved measures of integrated information.
\newblock {\em PLoS computational biology} {\bf 2016}, {\em 12},~e1005123.

\bibitem[Albantakis \em{et~al.}(2014)Albantakis, Hintze, Koch, Adami, and
  Tononi]{albantakis2014}
Albantakis, L.; Hintze, A.; Koch, C.; Adami, C.; Tononi, G.
\newblock Evolution of Integrated Causal Structures in Animats Exposed to
  Environments of Increasing Complexity.
\newblock {\em PLOS Computational Biology} {\bf 2014}, {\em 10},~1--19.
\newblock
  doi:{\changeurlcolor{black}\href{https://doi.org/10.1371/journal.pcbi.1003966}{\detokenize{10.1371/journal.pcbi.1003966}}}.

\bibitem[Hartmanis(1966)]{hartmanis1966algebraic}
Hartmanis, J.
\newblock {\em Algebraic structure theory of sequential machines (prentice-hall
  international series in applied mathematics)}; Prentice-Hall, Inc.,  1966.

\bibitem[Egri-Nagy and Nehaniv(2008)]{egri2008hierarchical}
Egri-Nagy, A.; Nehaniv, C.L.
\newblock Hierarchical coordinate systems for understanding complexity and its
  evolution, with applications to genetic regulatory networks.
\newblock {\em Artificial Life} {\bf 2008}, {\em 14},~299--312.

\bibitem[Egri-Nagy and Nehaniv(2015)]{egri2015computational}
Egri-Nagy, A.; Nehaniv, C.L.
\newblock Computational Holonomy Decomposition of Transformation Semigroups.
\newblock {\em arXiv preprint arXiv:1508.06345} {\bf 2015}.

\bibitem[Mayner \em{et~al.}(2018)Mayner, Marshall, Albantakis, Findlay,
  Marchman, and Tononi]{mayner2018pyphi}
Mayner, W.G.; Marshall, W.; Albantakis, L.; Findlay, G.; Marchman, R.; Tononi,
  G.
\newblock PyPhi: A toolbox for integrated information theory.
\newblock {\em PLoS computational biology} {\bf 2018}, {\em 14},~e1006343.

\bibitem[Karnaugh(1953)]{karnaugh1953map}
Karnaugh, M.
\newblock The map method for synthesis of combinational logic circuits.
\newblock {\em Transactions of the American Institute of Electrical Engineers,
  Part I: Communication and Electronics} {\bf 1953}, {\em 72},~593--599.

\bibitem[Ellis(2016)]{ellis2016can}
Ellis, G.
\newblock How can Physics Underlie the Mind.
\newblock {\em Springer2016} {\bf 2016}.

\bibitem[Auletta \em{et~al.}(2008)Auletta, Ellis, and Jaeger]{auletta2008top}
Auletta, G.; Ellis, G.F.; Jaeger, L.
\newblock Top-down causation by information control: from a philosophical
  problem to a scientific research programme.
\newblock {\em Journal of the Royal Society Interface} {\bf 2008}, {\em
  5},~1159--1172.

\end{thebibliography}



\end{document}